# Low loss QKD optical scheme for fast polarization encoding

A. Duplinskiy[1,2,*], V. Ustimchik[1,3], A. Kanapin[1,4], V. Kurochkin[1] and Y. Kurochkin[1]

[1]*Russian Quantum Center (RQC), Business Center «Ural», 100, Novaya st., Skolkovo, Moscow reg., 143025, Russian Federation*
[2]*Moscow Institute of Physics and Technology, 9 Institutskiy per., Dolgoprudny, Moscow Region, 141700, Russian Federation*
[3]*IRE RAS, Mokhovaya St, 11 – 7, Moscow, 125009, Russian Federation*
[4]*Lomonosov Moscow State University, GSP-1, Leninskie Gory, Moscow, 119991, Russian Federation*
*[*]a.duplinskiy@rqc.ru*

**Abstract:** We present a new optical scheme for BB84 protocol quantum key distribution (QKD). The proposed setup consists of a compact all-fiber polarization encoding optical scheme based on $LiNbO_3$ phase modulators, single laser source and two single-photon detectors. Optical scheme consists of standard telecommunication components and is suitable for both fiber and free-space quantum communication channels. Low losses (~2dB) in Bob's device increase both the key generation rate and distance limit. A new technique for solving polarization mode dispersion (PMD) issue in $LiNbO_3$ is implemented, allowing two crystals to neutralize the effect of each other. Several proof-of-concept experiments have been conducted at 10 MHz repetition frequency over 50 km of standard optical fiber under laboratory conditions and over 30 km of urban fiber with high losses (13dB), which is a link within a QKD network. To achieve this, calibration algorithms have been developed, allowing the system to work autonomously and making it promising for practical applications.

## 1. Introduction

QKD is a rapidly developing technology. It allows unconditional secrecy of sharing keys between distant users – the transmitter (Alice) and the receiver (Bob), the security being guaranteed by fundamental laws of quantum physics. In QKD, Alice and Bob generate a secret key by transmitting information encoded in the states of single photons or weak coherent pulses. The two most widely used encoding methods are phase coding, in which information is contained in the phase difference between two modes interfering with each other, and polarization coding, in which the information is carried by the state of polarization (SOP) [1].

Polarization encoding is common for free-space applications, since the atmosphere, unlike optical fibers, keeps the polarization stable. However, even polarization drift in a fiber quantum channel (QC) can be overcome with the help of polarization stabilization techniques. Moreover, this problem is not unique for polarization encoding schemes – usually phase encoding QKD setups also require SOP recovery since active modulators are polarization dependent [2,3].

Achieving quantum key production at competitive rates requires high frequency generation of polarization states, which is a challenge. Polarization controllers available on the market do not allow GHz state preparation frequencies, common for present-day QKD setups. The simplest fast QKD configuration for Alice utilizes four independent laser sources, one for each SOP required for BB84 protocol [4-7]. However, it appears to be hard to guarantee the indistinguishability of pulses emitted from different lasers, resulting in the system's vulnerability [15,16]. Bob as well needs to modify the SOP arriving at his station to select the measurement basis. Due to the same demand of high speed, it is usually done passively with the help of a beamsplitter (BS) [8-14]. The main drawback of this method is that the number of single-photon detectors (SPDs) increases by a factor of two – from two to four. This results in a higher quantum bit error rate (QBER) as the number of noise clicks rise. Moreover, SPDs are practically the most expensive part of QKD devices.

An alternative idea to use Pockels effect of fast electro-optical LiNbO$_3$ phase modulators for switching the polarization has been proposed with two main methods implementing this idea. The first one is based on the balanced interferometers [17,18]. Two orthogonal polarization components enter different arms of the interferometer with the help of a polarization beam splitter, after that one of the components experiences a phase shift induced by the modulator. As a result, two diagonal and two circular states can be generated. However, fiber Mach-Zendner interferometers are very sensitive and require phase stabilization [21], while interferometers with constant phase difference use free-space components that increase losses [17,18,22]. Jofre et al. proposed a different approach to switching the SOP with the help of phase modulators [19]. The transmitter based on this technique produces orthogonal states as polarization maintaining fiber is aligned at an angle of 45° directly to the $LiNbO_3$ crystal inside the modulator. A phase difference between orthogonal polarization components is produced, since modulation affects only one axis. The critical issue of this method is polarization mode dispersion (PMD) caused by the birefringence of the crystal. Suggested solutions, including polarization maintaining fiber (PMF) compensating patch cords [19] and Faraday mirrors [20,22] complicate the optical scheme (see sec. 4).

We present a simple configuration polarization encoding scheme based on $LiNbO_3$ phase modulators both in Alice's and Bob's devices, which use a single laser source and only two SPDs, while solving the PMD issue described above. Another advantage is that in contrast to the transmitter based on the phase modulator described in [19], there is no need to carry out any specific manipulation to align the PMF and the modulator's crystal, as a polarization controller or PMF patch cord spliced at an angle could be used (see sec. 2, 3). This significantly simplifies the technology and makes it possible to use modulators in regular configuration, which are available on the market. The result is a compact all-fiber system which consists of only standard telecommunication fiber components with low losses of about 2 dB on Bob's side.

A proof-of-concept experiment has been carried out at a 10 MHz laser pulses repetition frequency over 50 km of single-mode optical fiber. The system operates autonomously with the help of calibration algorithms, developed to set the polarization controllers' voltages. Once the QBER exceeds a threshold value set by the user, recalibration is applied automatically (see sec. 6, 7). The average QBER in our work is 2% with a sifted key rate of 0.5 Kbit/s. Furthermore, the system has been tested as a part of an urban QKD network [26]. Experimental demonstration confirms the suitability of the setup for practical applications. In addition, electronics can be upgraded to reach much higher pulse rates, as the scheme is limited only by the modulators' maximum frequency, which can reach 10-40 GHz.

## 2. Experimental setup

The basic setup is shown on Fig.1. Alice produces linearly polarized optical pulses using a 1550 nm laser source. The subsequent polarization controller (PC 1) is configured in such a way that the amplitudes of the field along the ordinary and extraordinary axes of the crystal inside the modulator (PM 1) are equal. This allows using the phase modulator for generating two pairs of orthogonal polarization states – "linear" and "circular" bases (see sec.3). The final element in Alice's apparatus is a variable optical attenuator that has two modes of operation – key sharing and system calibration. During the former, light is attenuated to 0.1 photon per pulse, while the latter is performed with stronger pulses (more than 1 photon per pulse, depending on the losses) to speed up the tuning procedure. As soon as the error rate reaches an acceptably low level, the attenuation is switched back to the key distribution regime value.

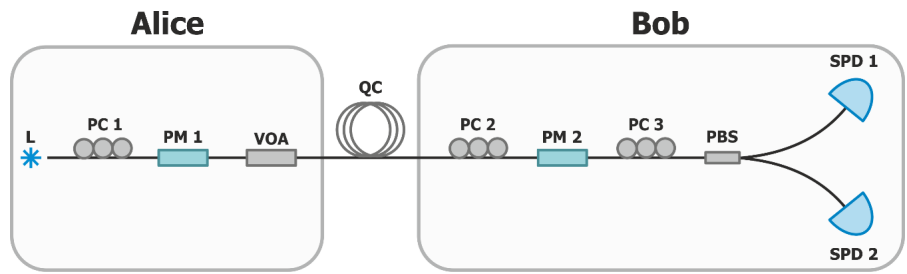

Fig. 1. QKD optical scheme for BB84 protocol with polarization encoding. Laser source (L) emits polarized optical pulses at 1550 nm. Polarization controller (PC 1) transforms the polarization state so that the amplitudes along the crystal axes of Alice's phase modulator (PM 1) are equal to each other. This allows Alice to encode

bits of the secret key in the SOP with the help of the modulator. To weaken the pulse, a variable optic attenuator (VOA) is used. The intensity is reduced to calibration or key generation level, depending on the operation mode. After the quantum channel (QC), the second piezo-driven polarization controller (PC 2) compensates SOP drifts and rotates it so that the polarization components along the lithium niobate crystal axes switch places, compensating the birefringence of $LiNbO_3$. Bob's modulator PM 2 is used for basis selection. Finally, polarization controller PC 3 converts SOPs for polarization beam splitter (PBS) to distinguish states with the help of single-photon detectors (SPD1, SPD2). Standard single-mode fiber is suitable for all elements included; however, three polarization controllers are used.

At the input of the Bob's device is a polarization controller (PC 2), that compensates the changes that SOP of the pulse undergoes within the QC. In addition, it is used to rotate the polarization components by 90° relative to their positions at the input of the Alice's modulator (PM 1). As a result, the component that passed along the fast axis of lithium niobate inside Alice's modulator travels along the slow axis of the identical crystal within Bob's modulator and vice versa, so the two $LiNbO_3$ crystals of the modulators compensate each other's polarization mode dispersion (sec. 4). Bob uses his modulator (PM 2) to select the measurement basis, deciding whether to apply a phase shift corresponding to a $\lambda/4$ plate, thereby switching between "linear" and "circular" bases (see sec. 3). The final polarization controller (PC 3) rotates the polarization, preparing the light for projective measurement in the corresponding basis with the use of a polarization beam splitter (PBS) and two single-photon detectors (SPD 1, SPD 2). The procedure is similar to using a half-wave plate in free-space (Fig. 2), rotating polarization by 45°, to fit our logical state polarizations into one of two PBS output channels depending on the state (sec. 3). For continuous operation, all polarization controllers are piezo-driven and are automatically adjusted by feedback algorithm described in the sec. 5.

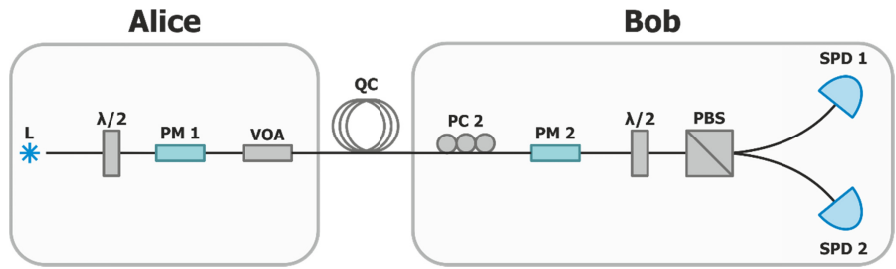

Fig. 2. A version of the optical scheme with free space elements and single polarization controller. Controllers PC1 and PC3 have been replaced with free-space optical elements – half-wave plates and PBS cube. Alice and Bob use PMF inside their devices.

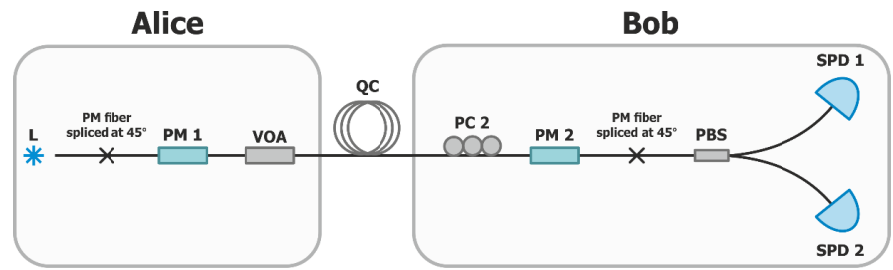

Fig. 3. Full-fiber version of the optical scheme with single polarization controller. Controllers PC1 and PC3 have been replaced with fiber splices at a 45° angle. Alice and Bob use PMF inside their devices.

Polarization controllers PC 1 and PC 3 only affect the fiber inside the devices of Alice and Bob, so they could be replaced with passive optical components and PMF. Half-wave plates rotate the SOP and PMF retains the amplitudes along both axes constant. However, using free-space optics is less practical, due to its sensitivity and high losses of fiber in-out interfaces. Therefore, the most promising design is replacing controllers with PMF fiber splices at 45°, which provides the same effect (Fig. 3). In addition, any combinations of scheme versions could be used. Detailed analysis of these solutions is given in the sections 3 and 5.

### 3. Phase modulator for polarization state changing

Polarization state at the input of phase modulator can be represented as a Jones vector. Here and below we use the reference frame associated with the modulator's crystal axes. Ignoring the constant phase, the polarization state vector at the input of the modulator is expressed as follows:

$$\left\| \begin{array}{c} A \\ Be^{i\varphi_1} \end{array} \right\|. \qquad (1)$$

Here A and B are the real amplitudes along the ordinary and extraordinary axes of lithium niobate and $\varphi_1$ is the phase difference between the components [23]. To describe the influence of the phase modulator on the polarization we use Jones matrix[1]:

$$\left\| \begin{array}{cc} e^{i\varphi_{or}} & 0 \\ 0 & e^{i(\varphi_{ex}+\Delta\varphi)} \end{array} \right\|. \qquad (2)$$

Here $\Delta\varphi$ is a voltage-induced phase shift and $\varphi_{or}$ and $\varphi_{ex}$ – phase shifts along ordinary and extraordinary axes, respectively, at zero voltage.

The phase shift $\Delta\varphi$ will affect only one of the vector components. In order for the phase modulation to switch SOP to an orthogonal one, it is necessary and sufficient that the magnitudes of the vector components along the axes are equal, while the applied phase shift is $\pi$ [23]:

$$|A| = |B|, \quad \Delta\varphi = \pi. \qquad (3)$$

In such a configuration the phase modulator acts like a $\lambda/2$ plate, reflecting the polarization state symmetrically relative to the crystal axis. Therefore applying 0 or $\pi$ phase shifts we produce a pair of states which forms the first basis for BB84 protocol[2]:

$$\begin{aligned} \psi_1 &= \frac{1}{\sqrt{2}}\left(|\leftrightarrow\rangle + e^{i\varphi_1}|\updownarrow\rangle\right) \\ \psi_2 &= \frac{1}{\sqrt{2}}\left(|\leftrightarrow\rangle + e^{i(\varphi_1+\pi)}|\updownarrow\rangle\right) \end{aligned} \qquad (4)$$

To create the second basis states, we add an extra $\pi/2$ phase shift (similarly to $\lambda/4$ plate) to both of the states listed above:

---

[1] For simplicity, we describe a system without losses, since fixed losses do not affect the polarization state but only weaken the pulse's intensity. Polarization dependent losses (PDL) are considered to be negligible, as confirmed by experimental results.

[2] At this step we neglect the birefringence of the LiNbO$_3$ as it would be discussed in detail in sec.4

$$\chi_1 = \frac{1}{\sqrt{2}}\left(|\leftrightarrow\rangle + e^{i(\varphi_1+\pi/2)}|\updownarrow\rangle\right)$$
$$\chi_2 = \frac{1}{\sqrt{2}}\left(|\leftrightarrow\rangle + e^{i(\varphi_1+3\pi/2)}|\updownarrow\rangle\right)$$

(5)

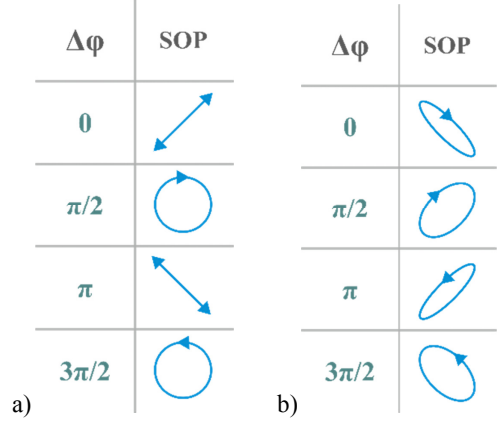

Fig. 4. Polarization patterns at the LiNbO$_3$ phase modulator output.
a) special case $\varphi_1 = 0$; b) arbitrary $\varphi_1$

In Fig.5, the set of corresponding SOPs is shown on the Poincare sphere. The four states are located on a meridian that includes diagonal and circular polarization states and are separated by 90° angles. The common offset angle is determined by $\varphi_1$. In the special case when $\varphi_1 = 0$, the first basis corresponds to linear diagonal states and the second to right and left circular polarizations (Fig.4a). For simplicity, we shall thereafter call the bases "linear" and "circular" even for nonzero $\varphi_1$.

Such a configuration of states is suitable for the BB84 protocol. That is, if Bob's apparatus is set to distinguish the states in one of the two bases, the states in the other basis will randomly produce an event in either detector with a probability ½.

Note that actually $\varphi_1$ does not have to be zero and in general case the states that fit our requirements are two pairs of orthogonal ellipses (Fig. 4b). This fact significantly simplifies the procedure of guiding the light into the modulator and is exploited within the experimental setup that we present.

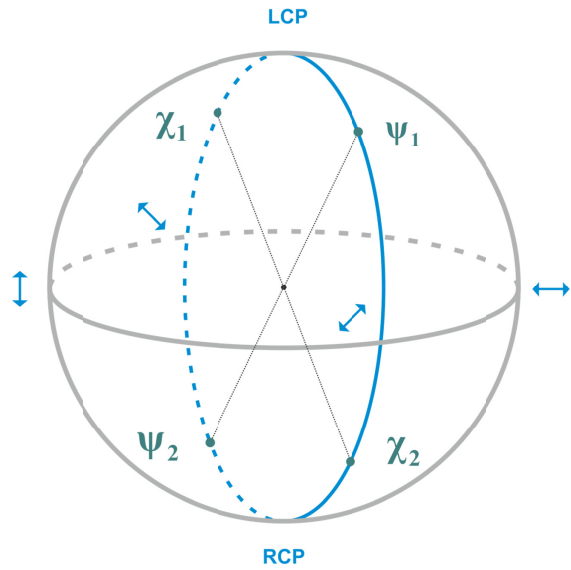

Fig. 5. The four states produced by Alice displayed on the Poincare sphere.

In order to fulfill condition (3), it is necessary to guide the light into the crystal in a way that the amplitudes along the crystal axes are equal. Similarly, for Bob to decipher the information correctly, an inverse transformation is required, turning the states into horizontal and vertical. As discussed above, this can be realized in free space by means of a half-wave plate, of in fibers via a polarization controller, or simply splicing two PMFs at a 45° angle.

## 4. Compensating polarization dispersion in a LiNbO$_3$ crystal

The lithium niobate crystal used in the phase modulators has a significant difference in the refraction indices of the ordinary and extraordinary axes. As a result, PMD is observed when light enters the crystal at an angle to its axes. The difference in the optical paths of the axes within a modulator is about 0.5 mm (less than 2 ps), so for pulses wider than a nanosecond the PMD itself does not produce significant error rate. However, power modulation of a semiconductor laser diode, which is used to generate the pulses, results in phase variation with time, i.e. in chirping of these pulses. This effect combined with a shift between the two orthogonal polarization components caused by the PMD leads to a significant degree of polarization degradation as SOP changes dramatically within a single pulse. This in turn causes a high error rate.

Several methods for compensating the occurring dispersion have been proposed. For example, PMF also possesses a difference in the refractive index of the two axes, so two modes could be aligned by choosing a patch cord of certain length [19]. Another alternative is guiding the pulse through the modulator a second time after being reflected from a Faraday mirror, so both components pass the ordinary and extraordinary axes one time each [20].

The solution presented in this work does not require any additional elements in the scheme. The polarization components entering Bob's phase modulator are rotated by 90° relative to the way they enter the modulator in the Alice's device. The component that passes through the ordinary axis in Alice's crystal passes through the extraordinary axis in Bob's crystal, and vice versa. As a result, the optical paths of both components are evened out, with the help of identical modulators and correct tuning of the polarization controller between them.

The absence of extra components minimizes losses in Bob's device, increasing the key bitrate. The calibration task for the polarization controller is discussed in detail in sec. 5.

## 5. The calibration procedure

The correct operation of the scheme requires the voltages of the piezo-driven polarization controllers to be set up in such a way that the polarization of the light entering both modulators and the PBS fits the requirements described above. Calibration is performed automatically. The input data for the tuning is the SPDs count statistics. To increase the amount of detector clicks and speed up the process, the light intensity is significantly increased for the duration of the calibration with the help of a voltage-driven attenuator. The intensity of the calibrating pulses sent by Alice is chosen depending on the losses in the quantum channel. Analyzing this data, the program adjusts the polarization controllers' voltages.

**Goal**

Let us briefly summarize the requirements for the SOPs described above:
1) At the input of Alice's phase modulator, the components along ordinary and extraordinary axes are equal
2) At the input of Bob's modulator, the components swap
3) Bob's measurements differentiate BB84 orthogonal states with extinction higher than 98%.

These requirements can be formulated mathematically by defining the Jones matrices for each of the scheme's sections.

The first section connects the laser source and Alice's phase modulator. We assume the incident light to be linearly polarized. To fulfill the criterion 1 the transformation of this section has to be:

$$\frac{\sqrt{2}}{2} \left\| \begin{matrix} e^{i\varphi_1} & 1 \\ 1 & e^{-i\varphi_1} \end{matrix} \right\| \quad (6)$$

The second section lies between Alice's and Bob's modulators and includes QC. According to the condition 2 the overall transformation of this part should be:

$$\left\| \begin{matrix} 0 & 1 \\ e^{i\varphi_2} & 0 \end{matrix} \right\| \quad (7)$$

Finally, the condition 3 is implemented by section between Bob's modulator and the PBS:

$$\frac{\sqrt{2}}{2} \left\| \begin{matrix} e^{i\varphi_3} & 1 \\ 1 & e^{-i\varphi_3} \end{matrix} \right\| \quad (8)$$

All three sections may introduce arbitrary relative phase shifts ($\varphi_1$, $\varphi_2$, $\varphi_3$) between the polarization components, but in order for PBS to distinguish the states correctly their sum should be divisible by $2\pi$:

$$(\varphi_1 + \varphi_2 + \varphi_3) \vdots 2\pi \quad (9)$$

This is required as production of the matrices (6), (7) and (8) should be equal to identity matrix. Note that $\varphi_1$ mentioned in sec. 3 is a result of the transformation (6) that is applied in the first section.

**Tuning**

As we cannot directly measure the elements of the Jones matrices for every section, we need to establish a set of observed values to implement the calibration. We tune one controller after another relying on the single photon detector counting statistics. To get enough information we need to switch on phase modulators and match the detector statistics to the phase shifts that we apply.

Alice applies four different voltages to the phase modulator, corresponding to phase shifts of 0, $\pi/2$, $\pi$ and $3\pi/2$. Bob applies two voltages, corresponding to 0 and $\pi/2$. These shifts do not mean correct polarization states production and detection, since polarization controllers are not tuned correctly yet. We begin the calibration by amassing the statistics of detector clicks corresponding to each pair of voltages, receiving a histogram of eight columns (Fig. 6).

The tuning of PC2 is based on the idea that, if correctly set up, the two phase modulators apply their phase shifts to orthogonal polarization components of light (sec. 4). Therefore, the pulses that experienced the pairs of shifts (0, 0) and ($\pi/2$, $\pi/2$) should not be distinguishable. This means that these two pulses will have the same statistics of clicks.

In addition, the following pairs are indistinguishable:
1) with a relative phase shift of $\pi/2$: ($\pi/2$, 0) and ($\pi$, $\pi/2$);
2) with relative phase shift of $\pi$: ($\pi$, 0) and ($3\pi/2$, $\pi/2$);
3) with relative phase shift of $3\pi/2$: ($3\pi/2$, 0) and (0, $\pi/2$).

Thus, we tune PC 2 aiming to equalize the click statistics for these combinations. Once PC 2 is set up correctly, PC 1 and PC 3 may be tuned in any order. The first one maximizes the difference between logical 1 and 0 in the correct bases, while the third one finishes the procedure, minimizing the error rate. The last controller that is tuned fits Eq. (11) equation for the current setups of the two controllers that were calibrated before.

The setup upgrades using passive elements (Fig. 2, Fig. 3) significantly simplifies the calibration process, as only one controller has to be tuned. Indeed, sections 1 and 3 are already consistent with matrices (6) and (8). The only polarization controller in the scheme has to minimize the error rate, which means that it meets the (7) type, also fulfilling the condition (9). Recalibration may be caused by the changes in the QC or by drifts of $\varphi_1$ or $\varphi_3$, that are not guaranteed to remain stable.

## Algorithm

Three channels of the polarization controllers are used. The controller successively minimizes the parameter for each channel using gradient descent.

Each controller uses its own parameter for minimization. Thus, for PC2, which attempts to align four pairs of columns on the histogram, the parameter was chosen to be equal to the squared sum of differences of corresponding columns, averaged over a short period to reduce the influence of noise on this parameter. The time has been chosen manually to ensure a balance between precision and performance speed. PC1 aims to achieve a maximal difference between logical 0 and 1 columns when Alice and Bob bases are the same. The parameter is the difference, taken with a minus sign. PC3 directly minimizes the error rate.

## 6. Experimental results

A proof-of-principle experiment has been conducted at a 10 MHz repetition frequency. The key distribution has been carried out over a distance of 50 km of standard single-mode optical fiber in spool (10 dB losses) with 0.1 photons per pulse. The system automatically performed the calibration procedure and maintained QBER below 5% for hours, applying recalibration as needed. Average time that the system has spent in the data transfer mode is about 80%, while the other 20% has been required for recalibrations. Unfortunately, the effective key generation frequency was reduced to 5 MHz due to the processing issues. The sifted key generation rate of 0.5 Kbit/s and 2% QBER have been obtained.

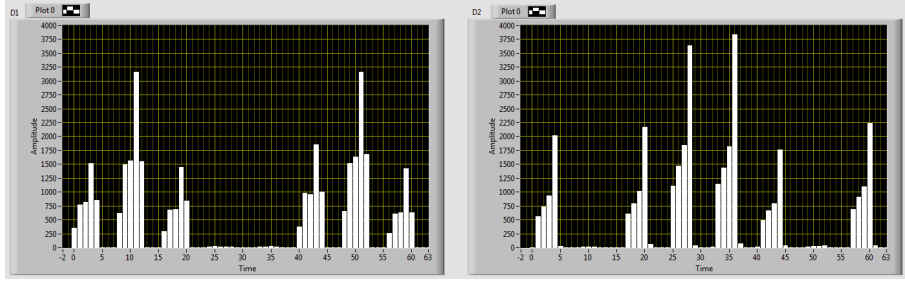

Fig. 6. Histogram of eight pulses for two detectors, illustrating all possible combinations of Alice's and Bob's phase shifts, corresponding to the table below.

Table 1. All possible combinations of Alice's and Bob's states. Letters "C" and "L" stand for "circular" and "linear" states, respectively. For more detail, see sec. 3.

| Pulse number | 1 | 2 | 3 | 4 | 5 | 6 | 7 | 8 |
|---|---|---|---|---|---|---|---|---|
| Alice's bit | 0 | 1 | 1 | 0 | 0 | 1 | 1 | 0 |
| Alice's basis | C | L | C | L | C | L | C | L |
| Bob's basis | L | L | L | L | C | C | C | C |
| Bob's bit after sifting | - | 1 | - | 0 | 0 | - | 1 | - |

In addition, an environment experiment has been carried out for a 30 km urban line with high losses (13 dB). To suppress the noise caused by the nearby fiber channels, Bob used a WDM optical filter at 1554.94 nm. The laser has been tuned with the help of temperature controller to fit the filter's wavelength. In order to guarantee the key secrecy, the average number of photons per pulse has been lowered to 0.02. Under these conditions, 106 bit/s sifted key rate has been obtained, taking calibration time into account. Fig. 8 illustrates QBER statistics vs. time during urban tests for 20 hours, the average value being 5.5%. The post-processing procedure applied to the sifted key consists of information reconciliation, parameter estimation, privacy amplification, and authentication check stages [25]. As a result, a secret key has been generated with 0.02 Kbit/s rate. This value could be improved in the future by using the decoy-state protocol [24], which allows increasing the intensity of the pulses. Upgrade of the driving electronics will provide higher repetition frequencies that will also significantly increase the secret key rate.

The system is currently used as a link within a QKD network across the urban fiber channels [26]. The developed QKD network is based on the trusted repeater paradigm and allows establishing a common key between users over an intermediate trustworthy node. The second link is a "plug and play" phase encoding scheme.

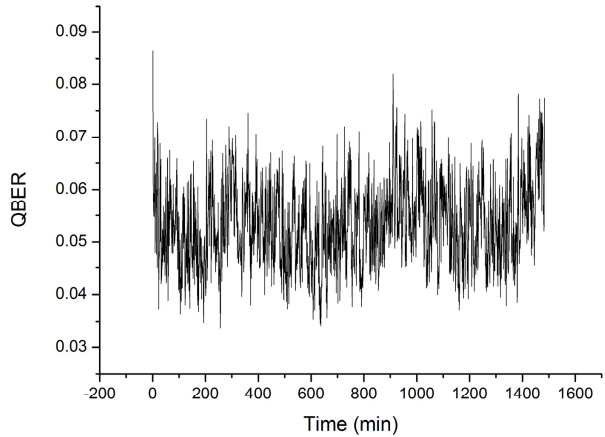

Fig. 7. QBER statistics with time during the urban tests with 13 dB losses and 0.02 photons per pulse.

The system uses ID Quantique ID230 single-photon detectors. Data acquisition, control of the phase modulators, attenuator, laser source and polarization controllers is carried out via a National Instruments FPGA, the software being written in LabVIEW.

## 7. Conclusions

A novel optical scheme implementing the polarization encoding BB84 protocol has been presented. Alice uses a $LiNbO_3$ phase modulator to generate two pairs of orthogonal polarization states with a single laser source, solving the issue of pulses' indistinguishability. A polarization controller or a PMF splice at an angle of 45° guide the pulses into the modulator with equal amplitudes along the crystal axes. Bob's device similarly selects the measurement basis with a modulator and rotates the output SOP in order for the PBS to distinguish different bits. Only two SPDs need to be used due to the active basis selection. Low losses (~ 2 dB) in Bob's apparatus allow increasing the communication distance and the bitrate. In addition, a novel approach solving the issue of PMD caused by the $LiNbO_3$ birefringence is implemented. A polarization controller following the quantum channel rotates the SOP by 90° equalizing the optical paths of two polarization modes within the crystals. A significant advantage of the scheme is that it consists of only standard telecommunication components and is suitable for both fiber and free space QCs.

Proof-of-concept experiments have been conducted at 10 MHz over 50 km of optical fiber in a spool and 30 km of urban fiber line. The system has been used as a part of a QKD network. To achieve continuous operation for several hours, calibration algorithms have been developed, allowing the system to work autonomously. The algorithm has proved itself reliable under both laboratory and urban conditions, spending approximately 20% of time for recalibration. The scheme seems suitable for future upgrades to higher frequencies and decoy-state protocols, being promising for practical applications.

## Acknowledgments

The research leading to these results has received funding from Russian Science Foundation under project 17-71-20146.